# Electrical switching of vortex core in a magnetic disk

Keisuke Yamada[1*], Shinya Kasai[1*], Yoshinobu Nakatani[2], Kensuke Kobayashi[1], Hiroshi Kohno[3], André Thiaville[4] & Teruo Ono[1]

[1]*Institute for Chemical Research, Kyoto Univeristy, Uji 611-0011, Japan*

[2]*University of Electro-communications, Chofu 182-8585, Japan*

[3]*Graduate School of Engineering Science, Osaka University, Toyonaka 560-8531, Japan*

[4]*Laboratoire de physique des solides, CNRS and Univ. Paris-Sud, 91405 Orsay, France*

*These authors contributed equally to this work.

**A magnetic vortex is a curling magnetic structure realized in a ferromagnetic disk, which is a promising candidate of a memory cell for future nonvolatile data storage devices [1]. Thus, understanding of the stability and dynamical behaviour of the magnetic vortex is a major requirement for developing magnetic data storage technology. Since the experimental proof of the existence of a nanometre-scale core with out-of-plane magnetisation in the magnetic vortex [2], the dynamics of a vortex has been investigated intensively [3-10]. However, the way to electrically control the core magnetisation, which is a key for constructing a vortex core memory, has been lacking. Here, we demonstrate the electrical switching of the core magnetisation by utilizing the current-driven resonant dynamics of the vortex; the core switching is triggered by a strong dynamic field which is produced locally by a rotational core motion at a high speed of several hundred m/s. Efficient switching of the vortex core without magnetic field application is achieved thanks to resonance. This opens up the potentiality of a simple magnetic disk as a building**

**block for spintronic devices like a memory cell where the bit data is stored as the direction of the nanometre-scale core magnetisation.**

The manipulation of magnetisation by spin currents is a key technology for future spintronics [11-15] since it would liberate us from applying the external magnetic field to control devices. The underlying physics is that spin currents apply a torque on the magnetic moment when the spin direction of the conduction electrons has a relative angle to the local magnetic moment. This leads us to the hypothesis that any type of spin structure with spatial variation, such as magnetic vortex, can be excited by a spin-polarized current in a ferromagnet. Indeed, we have succeeded in showing that a magnetic vortex core can be resonantly put into a steady circular motion by an AC current flowing through a ferromagnetic disk, when the current frequency is tuned to the eigenfrequency originating from the confinement of the vortex core in the disk [10]. Through the spin-transfer effect, the current exerts a transverse force on the vortex core; on resonance, it amplifies a circular motion of the core, which is eventually balanced by the damping force, and a steady circular motion is realized. Beyond that, we discovered that higher excitation currents induce even the switching of the core magnetisation during the circular motion (3D movie of micromagnetic simulation is available as "Supplementary Information"), which is at the heart of the present study.

Prior to the experimental confirmation, we clarify the mechanism of the electrical core switching based on the results of micromagnetic simulations for the dynamics of the magnetic vortex. The current-induced dynamics of a vortex was calculated by micromagnetic simulations in the framework of the Landau-Lifshitz-Gilbert (LLG) equation with a spin-transfer term [10, 16, 17]. In the simulations, the electric current in the disk was assumed to be uniform in space, and oscillating in time as $j = J_0 \cos(2\pi f t)$. In the following, we present the calculation results for a disk with thickness $h$ = 50 nm and radius $R$ = 500 nm, which has the same dimensions as the sample used in the present experiments. The AC frequency is set to $f$ = 370 MHz, which is the resonance frequency of this disk with standard material parameters and nominal dimensions.

Figures 1a–f are successive snapshots of the calculated results for the magnetisation distribution during the process of core motion and switching, showing that the reversal of the core magnetisation takes place in the course of the circular motion without going out of the disk. Noteworthy is the development of an out-of-plane magnetisation (dip) which is opposite to the core magnetisation (Figs. 1a-d). This structure is different from the static ring-shaped dent due to the core stray field (Fig. 1a) in that it is much deeper and located only on one side of the core. This structure has already been noted in the context of field-driven dynamics [6] and can be understood as follows. It is known since the seminal work of W. Döring [18, 19] that magnetic structures in stationary motion at velocity $\vec{v}$ can be obtained by minimizing a kinetic potential given by the usual micromagnetic energy, without external fields but with the addition of a kinetic term. The effective field derived from this kinetic term (that creates the so-called gyrotropic force) is expressed as

$$\vec{H}_{kin} = \frac{1}{\gamma_0} \vec{m} \times \left[ (\vec{v} \cdot \vec{\nabla}) \vec{m} \right], \qquad (1)$$

where $\gamma_0$ is the gyromagnetic ratio and $\vec{m}$ is the local magnetisation unit vector. For a vortex moving along the in-plane $x$ direction in an infinite sample, this field has an out-of-plane *z* component localized on the *y* sides of the core, with both signs, and a maximum value given, for the static core profile in the zero film thickness limit [20], by

$$H^z_{kin} = 0.69 \frac{v}{\gamma_0 \Lambda}, \qquad (2)$$

with $\Lambda$ the micromagnetic exchange length (5 nm for permalloy). While the core motion is not linear but circular in the disk, the above analysis is still valid because the radius of the circular motion is much larger than the core size. The rotation sense depends on the core out-of-plane magnetization [8], and the dip appears always on the inner side of the core in agreement with Fig.1. As the core is accelerated, the dynamic field becomes larger and the dip is deepened (Figs. 1c-d), eventually approaching $m_z = -1$ (Fig. 1e). This is the limit of the dynamic deformation of the magnetic structure, and we can remark that large exchange energy is stored in this structure. The vortex core then reverses. In fact, the simulation shows that the dip build-up is suddenly accelerated from $m_z \approx -0.5$ to $m_z = -1$ in the last nanosecond (Figs. 1b-e), and that the core reversal starts as soon as the value $m_z = -1$ is reached. We have confirmed by three-dimensional micromagnetic simulations on smaller disks that a Bloch point, a singularity of the magnetisation, performs the core reversal as in the case of the field-driven switching of the core [21, 22], and as dictated by topology. This is followed by the merging of the two regions with negative $m_z$, leaving behind a negative-$m_z$ core vortex (Fig. 1f).

Figure 2a displays the vortex core trajectory corresponding to Fig.1. One sees that, after switching, the core steps towards the disk centre and starts rotating in the opposite sense, again evidencing the core switching. Figure 2b shows the in-plane velocity of the core as a function of time. The sudden decreases of velocity correspond to the repeated core-switching events. Worth noting is that the core switches when its velocity reaches a certain value, $v_{switch} \approx 250$ m/s here, regardless of the value of the excitation current density. The existence of a maximum velocity results directly from the Döring's analysis, as expressed by the equation (1). From the equation (2), one realizes that this velocity corresponds to a very large out-of-plane field $\mu_0 H_{kin}^z \approx 0.2$ T (in the zero-thickness limit), as indeed required for tilting the magnetisation out of the plane. While this value falls in the same range as the static homogeneous perpendicular field to reverse the core magnetisation [21], the present core reversal is caused by the local dynamic field induced by the resonant motion of the vortex, which is a kind of a breakdown similar to the Walker breakdown for a moving 1D Bloch wall [23, 24] (see Supplementary Discussion).

We confirmed the predicted current-induced switching of the vortex core by the magnetic force microscopy (MFM) observation as described below. Figure 3a shows an atomic force microscopy (AFM) image of our sample fabricated on the thermally oxidized Si substrates by the lift-off method in combination with e-beam lithography. The sample has a Permalloy disk with the thickness and the radius of 50 nm and 500 nm, respectively. The disk is sandwiched between two wide Au electrodes with 50nm-thickness, through which an AC excitation current is supplied (also see the scanning electron microscope image in the inset of Fig.4).

First, the direction of a core magnetisation was determined by MFM observation. A dark spot at the centre of the disk in Fig. 3b indicates that the core magnetisation directs upward with respect to the paper plane. The core direction was checked again after the application of an AC excitation current of a frequency $f = 290$ MHz and an amplitude $J_0 = 3.5 \times 10^{11}$ A/m² through the disk with the duration of about 10 sec. Here, the current densities were evaluated by dividing current by the disk diameter times thickness. As shown in Fig. 3c, the dark spot at the centre of the disk changed into bright after the application of the excitation current, indicating that the core magnetisation has been switched. Figures 3b to 3l are successive MFM images with an excitation current applied between each consecutive image. It was observed that the direction of the core magnetisation after application of the excitation current was changed randomly. This indicates that the switching occurred frequently compared to the duration of the excitation current (about 10 sec.) and the core direction was determined at the last moment when we turned off the excitation current.

The experiment described above was performed for various frequencies of the excitation current. The experiment was repeated 20 times for each current frequency and the number of core switching times was counted for each frequency. Figure 4 shows the switching probability as a function of the frequency of excitation current. Here, the switching probability is defined as the ratio of the number of the core switching events to the total trial number (20 times). The switching probability is high in a certain region of excitation current frequency, corresponding to the resonance frequency of the core motion in the disk. As mentioned above, the switching probability of one half corresponds to the core switching with much faster time scale than the duration of the excitation current. Thus, the switching probability should be half for a wider frequency range if we apply higher excitation current. As $J_0 = 3.5 \times 10^{11}$ A/m$^2$ is the maximum current density in our present experimental setup, this could not be checked. However, by decreasing the current to $2.4 \times 10^{11}$ A/m$^2$, the probability peak shrinks as expected. In Fig.4, we also superpose to the experimental data the results of micromagnetic simulations. They show a similar frequency range for switching for a current about 40 % above the threshold, and the minimum current densities for switching are quite close (2.8 vs. 2.4 × $10^{11}$ A/m$^2$). It should also be noted that the resonance behaviour in Fig. 4 excludes the possibility of the core switching induced by Joule heating due to the excitation current, because the electric power is independent of the excitation frequency.

Thus, we have shown that an electric current can switch the direction of a vortex core in a magnetic disk. The current necessary for the switching is only several mA, while the core switching by an external magnetic field needs a large magnetic field of several kOe [21]. Although we demonstrated the repeated vortex core switching by a continuous AC current, it will be possible to control the core direction by a current with an appropriate waveform. The switching mechanism of the core on resonance discussed here also applies to the AC field driven case that was demonstrated independently very recently[25] (see Supplementary Discussion). The quantitative agreement between experiment and micromagnetic simulation demonstrates the validity of the micromagnetic description of the spin-transfer torque. Given the presently observed discrepancies between measured and calculated results on the current-induced domain wall motion, the agreement obtained here, on a non-trivial physical effect, is very important. The current-induced vortex core switching demonstrated here can be used as an efficient data writing method for a memory device in which the data is stored in a nanometre size core.

**Methods**

**Details of the calculations**

The current-induced dynamics of a vortex was calculated by micromagnetic simulations in the framework of the Landau-Lifshitz-Gilbert (LLG) equation with a spin-transfer term [10, 16, 17]. In the simulations, the electric current in the disk was assumed to be uniform in space, and oscillating in time as $j = J_0 \cos(2\pi f t)$. The assumption on the uniformity is justified by the fact that the radius of the core motion is smaller than the width of the electrode of the sample in the experiments. For the large disk ($h$ = 50 nm, $R$ = 500 nm), we performed the two-dimensional calculation by dividing the disk into rectangular prisms of $4 \times 4 \times 50$ nm$^3$; the magnetisation was assumed to be constant in each prism. We also performed the three-dimensional calculation for smaller disks ($h$ = 20, 40, 60 nm, $R$ = 120 nm) by dividing them into rectangular prisms of $4 \times 4 \times 5$ nm$^3$ to check the validity of the two-dimensional calculation. We carefully checked that the results show no appreciable change if we include in the simulation the magnetic (Oersted) field generated by the current. The typical material parameters for Permalloy were used: saturation magnetisation $M_s$ = 1 T, exchange stiffness constant $A = 1.0 \times 10^{-11}$ J/m, and damping constant $\alpha$ = 0.01. The spin polarization of the current was taken as $P = 0.7$ [26]. The non-adiabatic term [17] in the spin torque was set to $\beta$=0 here, as we checked by simulations that it did not have any appreciable influence on the results for 0<$\beta$<0.1, and as also understood from the analytical model (see Supplementary Discussion).

**Resonance frequency**

As the resonance occurs at $f$ = 290 MHz in the experiment, the calculated frequencies were rescaled by a factor 0.776 to match the observation. Such a discrepancy in the resonance frequency between the simulation and the experiment was often reported and the reduced magnetisation has been used for the simulation [6, 7]. Here, we just rescale the frequency instead of reducing the magnetisation in the simulation.

**Sample temperature**

By measuring the sample resistance, we estimated that the temperature rise of the sample by the application of the excitation current is only 10 K, which also excludes the possibility of the critical role of the Joule heating on the vortex core switching.

**Acknowledgements** The present work was partly supported by MEXT Grants-in-Aid for Scientific Research in Priority Areas and JSPS Grants-in-Aid for Scientific Research.

**Competing interests statement** The authors declare that they have no competing financial interests.

**Correspondence** and requests for materials should be addressed to T. O. (ono@scl.kyoto-u.ac.jp).

Figure 1 Perspective view of the magnetisation with a moving vortex structure. The height is proportional to the out-of-plane (z) magnetisation component. Rainbow colour indicates the in-plane component as exemplified by the white arrows in (a). (a) Initially, a vortex core magnetized upward is at rest at the disk

centre. (b) On application of the AC current, the core starts to make a circular orbital motion around the disk centre. There appears a region with downward magnetisation (called 'dip' here) on the inner side of the core. (c,d,e) The dip grows slowly as the core is accelerated. When the dip reaches the minimum, the reversal of the initial core starts. (f) After the completion of the reversal, the stored exchange energy is released to a substantial amount of spin waves. A positive 'hump' then starts to build up, which will trigger the next reversal. Calculation with $h$= 50 nm, $R$= 500 nm and $J_0$= 4 ×10$^{11}$ A/m$^2$. 3D movie of micromagnetic simulation is available as "Supplementary Information".

Figure 2 Core dynamics under AC spin-polarized current. The calculation was performed for the same disk as in Fig.1. (a) In-plane trajectory of the core motion of Fig.1 ($J_0$= 4 ×10$^{11}$ A/m$^2$), from $t$ = 17.5 to 23 ns with the snapshot points indicated. (b) Magnitude of the core velocity as a function of time, for two choices of the current density, showing that the maximum velocity is the same. Since the core reversal is followed by the emission of many spin waves and chaotic-looking magnetisation dynamics around the core (see Fig. 1f), the switching time shows large variations.

Figure 3 MFM observation of electrical switching of vortex core. (a) AFM image of the sample. A Permalloy disk fills inside the white circle. The thickness of the disk is 50 nm, and the radius is 500 nm. Two wide Au electrodes with 50nm-thickness, through which an AC excitation current is supplied, are also seen. (b) MFM image before the application of the excitation current. A dark spot at the centre of the disk (inside the red circle) indicates that the core magnetisation directs upward with respect to the paper plane. (c) MFM image after the application of the AC excitation current at a frequency $f$ = 290 MHz and amplitude $J_0$ = 3.5 × 10$^{11}$A/m$^2$ through the disk with the duration of about 10 sec.

The dark spot at the centre of the disk in Fig. 3b changed into the bright spot, indicating the switching of the core magnetisation from up to down. (b-l) Successive MFM images with excitation current applied similarly between consecutive images. The switching of the core magnetisation occurs from b to c, from f to g, from h to i, from i to j, and from k to l. The MFM scan area was limited by the experimental time, and some part of the sample which is not necessary to determine the core direction was omitted in the image. The low MFM contrast for the core is due to the height difference between the disk and the electrodes.

Figure 4 Switching probabilities as a function of excitation current frequency. The big symbols represent the experimental data, where the blue and red ones correspond to the case for $J_0$ = 2.4× $10^{11}$ A/m$^2$ and $J_0$ = 3.5 × $10^{11}$ A/m$^2$, respectively. The small green symbols are the simulation results with 100 ns duration and $J_0$ = 3.88 × $10^{11}$ A/m$^2$ (about 40 % above the calculated switching threshold of 2.8 × $10^{11}$ A/m$^2$), which corresponds to the larger experimental current. In plotting the simulation results, the frequencies were rescaled by a factor 0.776. (Inset) Scanning electron microscope image of the sample.

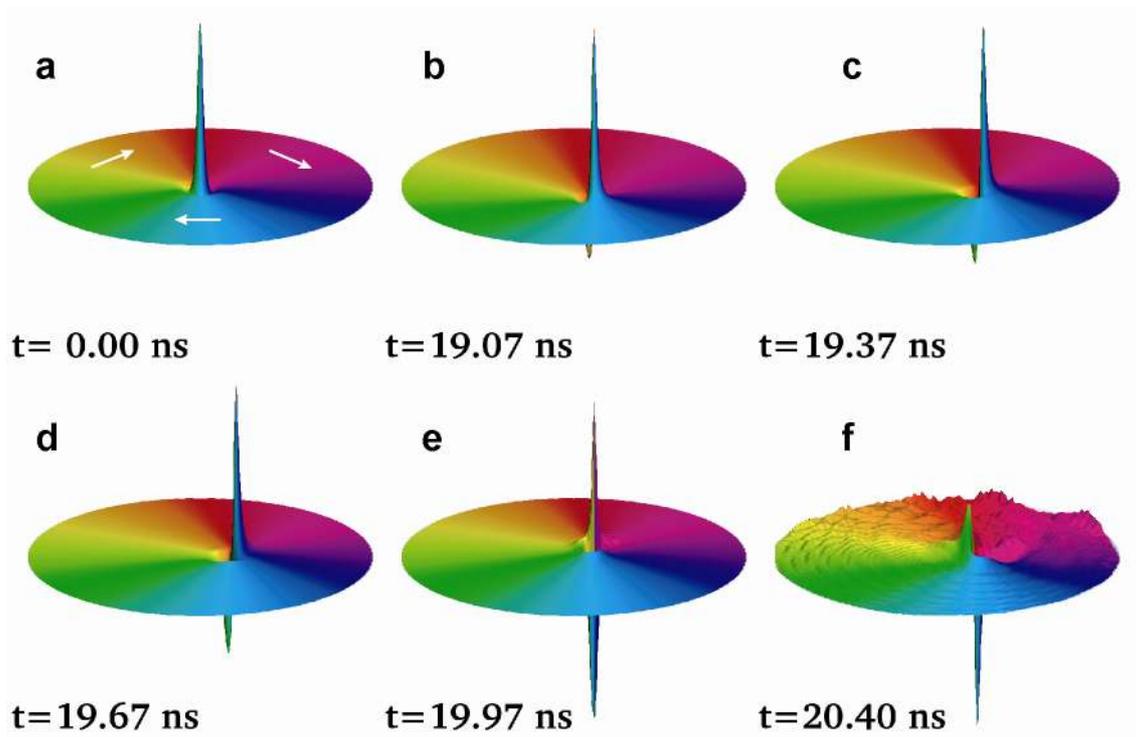

Figure 1

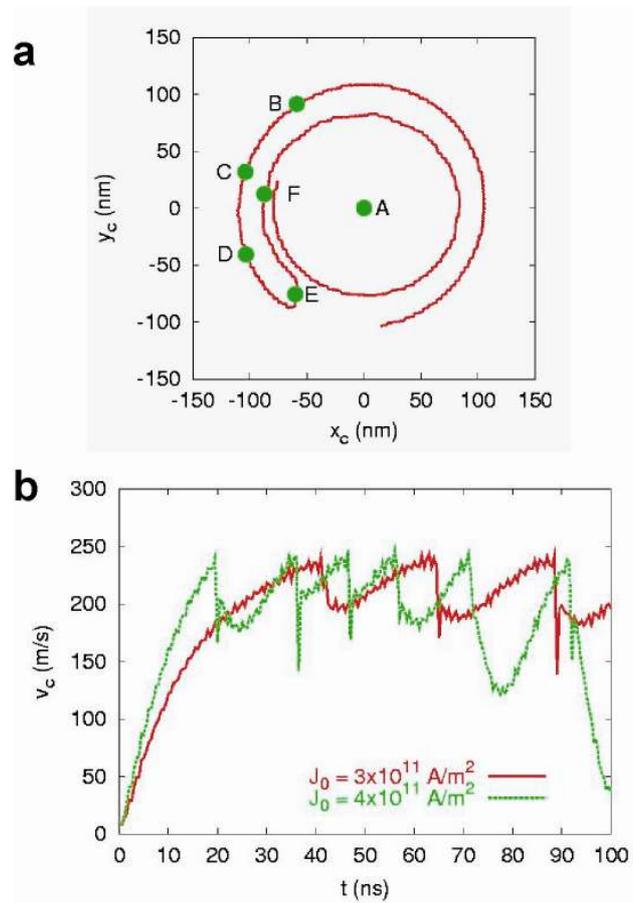

Figure 2:

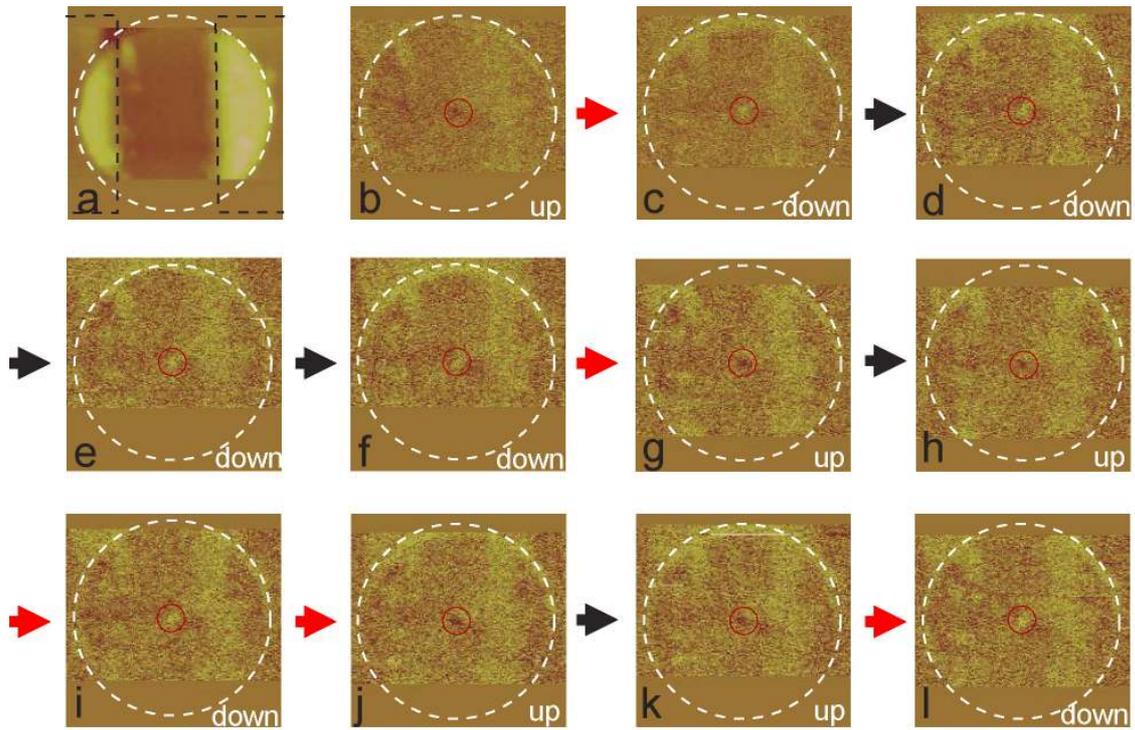

Figure 3

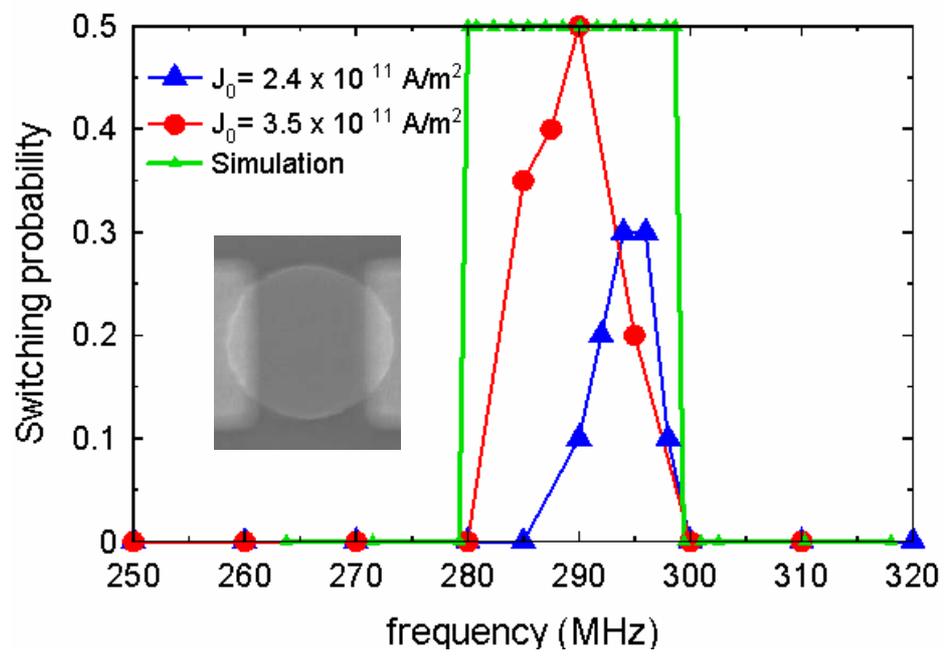

Figure 4

**Supplementary Discussion**

**Breakdown of the vortex core induced by the spin-polarized current**

We pointed out that the core reversal induced by the current is different from that caused by the application of the static magnetic field opposing to the core magnetisation. In fact, the present reversal is a breakdown, similar to the Walker breakdown for a moving 1D Bloch wall or, more closely, to the Bloch-line model for a twisted Bloch wall [22]. This means that, whereas small core velocities can be accomodated by a steady-state structure deformation (including the prominent dip), above the maximum velocity this is no longer possible and the structure continuously changes, here by reversing the core. One important difference with the Bloch-line model is the necessity of a micromagnetic singularity for the present breakdown. Therefore, one might expect that the switching current, velocity etc. depend a lot on the mesh size, similarly to the static and uniform field case [21]. However, since the appearance of the Bloch point is preceded here by the dip growth, which occurs continuously and not at the core, it is expected to be almost independent of mesh size. This was indeed checked in our micromagnetic simulations with decreasing mesh sizes. One can even get an order of magnitude estimation of the core velocity necessary for switching by writing

$$H_{kin}^z = M_s N_z (W_c / h), \quad (S1)$$

with $N_z$ the axial demagnetizing factor of a cylinder of height $h$ and diameter $W_c$ the vortex core diameter, that in the zero-thickness limit is $W_c = 2.64\ \Lambda$. Thus, the main physical parameter controlling the vortex core switching will be the core motion velocity.

**Core motion velocity in the resonance**

The core velocity becomes large here because of the resonance effect [10], and is only limited by the damping. This can be seen analytically using the point core model [8]. The Thiele force equation, generalized to include the effect of spin polarized current with the non-adiabatic term [8] gives, in the situation of an AC current, a steady-sate core motion described by

$$z_c(\omega) = i\frac{u_0}{2}\left(\frac{G - iD\beta}{\kappa + \omega G - i\alpha\omega D}\right)e^{-i\omega t} + \{\omega \to -\omega\}. \quad (S2)$$

In this formula, $z_c = x_c + iy_c$ is the complex core position, $u_0$ the velocity corresponding to the current $J_0$ and $\beta$ ($<<1$) the non-adiabaticity parameter of the spin transfer torque [17], $G$ the gyrovector $z$ component, $D$ the dissipation matrix element along the core trajectory, $\kappa$ the spring constant of the core-restoring force, and $\omega = 2\pi f$ is the pulsation [8]. The linearly polarized current is decomposed into two counter-rotating vectors, only one of which is resonant depending on the gyrovector sign ($G = \pm 2\pi\mu_0 M_s / \gamma_0$, the sign being that of the core out-of-plane magnetization). The resonant component produces a large core velocity at resonance

$$v_{c,res} = \frac{u_0}{2\alpha}\sqrt{\left(\frac{G}{D}\right)^2 + \beta^2} \approx \frac{u_0}{2\alpha}\frac{|G|}{D}. \quad (S3)$$

The value of $G$ is quantized [8] hence does not change, but the dissipation $D$ depends mainly on disk radius and also increases with core velocity, making (S3) nonlinear. To be more specific, the zero-thickness model gives for $R = 500$ nm $|G|/D = 2/5.7$ at rest, the micromagnetic computations for $R = 500$ nm and $h = 50$ nm give $|G|/D = 2/5.11$ at rest and $1/3$ close to switching. Thus, the core velocity becomes 16 times larger than the current equivalent velocity $u_0$. The negligible influence of the non adiabatic β term seen in the micromagnetic calculations is also simply understood.

The same model can be applied to the AC field driven case $h = H_0 \cos(2\pi f t)$. A simple estimate of the applied field energy in a square sample of edge $L$ shows that, in (S2), the numerator $u_0(G - iD\beta)$ has to be replaced by $\pm H_0 \mu_0 M_s h L$, where the sign reflects the circulation direction of the magnetization in-plane component. As a consequence, the same dynamics is obtained for $u_0 = H_0 \gamma_0 L / 2\pi$ (for example, our measured $u_0 = 15$ m/s translates into $\mu_0 H_0 = 0.4$ mT for the same damping, or 1 mT if α=0.05 as assumed[25] instead of our value α=0.02). Thus, our physical explanation of the core switching is general, and also shows that current-driven excitation becomes easier at small sizes.